\documentclass[preprint,showpacs,preprintnumbers]{revtex4}

\usepackage{graphicx}
\usepackage{dcolumn}
\usepackage{bm}
\usepackage{amsmath}

\begin{document}

\title{Renner-Teller effects in HCO$^{+}$ dissociative recombination}
\date{\today}

\begin{abstract}
A theoretical description of the dissociative recombination process for the
HCO$^{+}$ ion suggests that the nonadiabatic Renner-Teller coupling between electronic
and vibrational degrees of freedom plays an important role. This finding is
consistent with a recent study of this process for another closed-shell
molecule, the H$_{3}^{+}$ ion, where Jahn-Teller coupling was shown to
generate a relatively high rate. The cross section obtained here
for the dissociative recombination of HCO$^{+}$ exhibits encouraging
agreement with a merged-beam experiment.
\end{abstract}

\author{Ivan A. Mikhailov$^\dag$, Viatcheslav Kokoouline$^\dag$, {\AA}sa Larson$^*$, Stefano Tonzani$^\ddag$, Chris H. Greene$^\ddag$}
\affiliation{$^\dag$Department of Physics, University of Central Florida, Orlando, Florida 32816, USA; \\
$^*$ Theoretical Chemistry, School of Biotechnology, Royal Institute of Technology, AlbaNova University Center, S-106 91 Stockholm, Sweden; \\
$^\ddag$Department of Physics and JILA, University of Colorado, Boulder, Colorado 80309, USA}


\maketitle

\section{Introduction}

The dissociative recombination (DR) of small molecular ions that collide
with electrons plays an important role in interstellar diffuse and dense
clouds. It is well known that these clouds constitute building material for
new stars. 
The importance of DR cross sections as parameters in star formation models, for such astrophysically relevant ions as H$%
_{3}^{+}$ and HCO$^{+}$, is one of the reasons why DR has been extensively studied in laboratory
experiments \cite{larsson00}.
For diatomic
molecules the process is well understood and described theoretically \cite{hickman87,giusti80,giusti83,giusti84,nakashima87,jungen97,schneider97}. Until recently \cite{orel93,kokoouline01,kokoouline03a,kokoouline03b} though, theory was unable
to model the DR process in triatomic ions, except in cases where a neutral
dissociative state crossed the ionic Born-Oppenheimer surface in the
Franck-Condon region, generating a rapid rate \cite{orel93}. In
triatomics, a key complication is the fact that multiple vibrational  and
rotational  degrees of freedom must generally be taken into account. \ And
in addition to the greater computational burden of treating more dimensions
quantum mechanically, the addition of these new degrees of freedom can also
lead to new conceptual issues, related to the degeneracy of vibrational or
rovibrational levels in certain triatomic ions (e.g., in both H$_{3}^{+}$
and HCO$^{+}$). This can cause an intrinsic instability of the corresponding
neutral molecules \cite{LandauLifshitz} that causes them to distort away
from the symmetric geometry. 

Consider an incident electron that interacts with a closed-shell triatomic
ion having a degenerate vibrational mode. If the symmetry group $\Gamma $ of the resulting neutral complex has at least one 
degenerate irreducible representation, the electronic partial wave 
components with angular momentum $l>0$ typically contain at least one such representation,
whereby the corresponding electronic states of the neutral molecule are degenerate in
the clamped-nuclei approximation. Due to the Jahn-Teller theorem \cite{LandauLifshitz,jahnteller,longuet61}, first formulated by Landau 
\cite{LandauLifshitz,ballhausen79}, this degenerate
electronic state strongly interacts with a degenerate vibrational state of
the $\Gamma $ group. The interaction leads to either a quadratic (Renner-Teller) \cite{jungen80,duxbury98,mccurdy03} or linear conical
(Jahn-Teller) intersection, where ``quadratic'' or ``linear'' refers to the dependence of the adiabatic eigenvalues on the symmetry distortion coordinate; e.g., the electronically-degenerate linear molecule being considered in the present article displays a quadratic behavior near the symmetry point, as a function of the bending angle $\theta$.  
In electron-induced reactions, electron capture can be
followed by dissociation or autoionization of the recombined system. (In some rare cases, recombination could also occur radiatively following electron capture, but this is rarely important in molecular systems having an open dissociation channel and it will be ignored throughout this study.) Since
the degenerate electronic state causes the instability of the symmetric
configuration, the recombined molecule quickly distorts away from the
symmetry point to remove the degeneracy, after which the autoionization
channel typically becomes energetically closed, and the molecule eventually
finds a pathway along which it can dissociate (provided there is an electronic 
state of the molecule that is open for dissociation).
 In a system for which no
direct crossing of a neutral dissociative state with the ground state of the
ion is present, this indirect mechanism can become dominant, and it can still produce
large DR cross sections.  This is expected to be particularly true for
molecules containing hydrogen. 

Work by some members of the present collaboration have recently found that this is what happens in H$_{3}^{+}+e^{-}$
collisions \cite{kokoouline03b}. For HCO$^{+}$, a closed-shell linear ion in its ground state, the picture appears to be
similar. The lowest doubly-degenerate vibrational $E$ states are coupled
to the electronic states $E_{1}(np\pi )$ of the neutral system through the
Renner-Teller interaction, resulting in a large probability for recombination.

Although Renner-Teller coupling is a well-known phenomenon in spectroscopy, the
effect of Renner-Teller coupling on electron-molecule collisions has not been
studied until recently. McCurdy {\it et al.} \cite{mccurdy03} have made an
important contribution to understanding the role of Renner-Teller physics in
collisions between an electron and the CO$_2$ molecule. In a completely $ab$
$initio$ treatment, the authors employed a time-dependent framework to describe the nuclear motion of the negative ion CO$_2^-$ formed during the collision. The nuclei move on the two-component $^2A_1$ and $^2B_1$ electronic complex potentials corresponding to the doubly-degenerate $^2\Pi_u$ molecular state at linear configuration. The two components interact through the Renner-Teller coupling mechanism. 

The aim of the present study is to understand the mechanism of DR in HCO$%
^{+}$ and, in particular, to interpret its large measured rate. This
ion was first detected in space and then synthesized in the laboratory \cite{herbst74}.
Experimental measurements of DR in HCO$^{+}$ were realized in a
number of merged-beam, afterglow-plasma, and storage ring experiments \cite{lepadellec97,poterya05,adams84,amano90,ganguli88,gougousi97,laube98,leu73,rowe92,geppert05}. In Ref. \cite{larson05} we have presented the
potential energy surfaces for the ground molecular state of HCO$^{+}$ and the lowest
states of HCO up to principal quantum number $n=4$, calculated as functions of all three vibrational
coordinates. The potentials were then used to estimate the DR
rate. Since the splitting between the two $E_{1}(np\pi )$ electronic states
for the principal quantum number $n=3$ was found to be small, it was
suggested that the Renner-Teller non-Born-Oppenheimer coupling should not
play a significant role in DR of HCO$^{+}$, therefore, it was not included in
the model of Ref. \cite{larson05}. However, since even a small coupling
between degenerate states can in principle result in a large DR probability,
we investigate here the role of Renner-Teller coupling in HCO in
more detail, and find that we must revise that initial expectation.

The article is organized in the following way. In section \ref{sec:ham_RT} we construct the model Hamiltonian from the results of the {\it ab initio} calculation. The Hamiltonian includes the nonadiabatic Renner-Teller coupling and depends parametrically on the nuclear configurations. Section \ref{sec:cs} describes the quantum defect approach to obtain the reaction matrix from the Hamiltonian. It also describes how the DR cross section is obtained from the reaction matrix. Section  \ref{sec:results} presents and discusses the results of the calculation: the DR cross section and the thermally averaged DR rate. Section \ref{sec:auoionization} considers autoionization in detail. Autoionization is a process that competes with DR and it decreases the DR cross section. Section \ref{sec:stefano_paper} compares the present results with a previous theoretical study \cite{larson05}. Finally, section \ref{sec:conclusion} discusses our conclusions.

\section{Model Hamiltonian including the Renner-Teller coupling}
\label{sec:ham_RT}

The starting point of the present theoretical approach is the molecular
Hamiltonian $H$ of HCO, which  we represent as 
$H=H_{ion}+T_e+H_{int}$ ,
where $H_{ion}$ is the ionic Hamiltonian, $T_e$ is the kinetic energy of the incident electron, 
and $H_{int}$ describes the
electron-ion interactions. We assume that the ion is in its ground
electronic state. $H_{ion}$ and $H_{int}$ depend on the four internuclear
coordinates $\mathcal{Q}=\{R_{CH},R_{CO},\theta ,\varphi \}$, where $R_{CH}$
and $R_{CO}$ are the C-H and C-O internuclear distances, $%
\theta $ is the bending angle, which is zero for linear configurations. The
electronic energies are independent of the angle $\varphi $, which represents the azimuthal orientation of the bending. 

We assume that the incident electron is initially captured in one of the neutral Rydberg states, which can be
approximately characterized by the principal quantum number $n$, the orbital
angular momentum $l$, and its projection on the molecular axis $\lambda $.
In our model we have included only
the $np\pi ^{-1}$,$np\pi ^{+1}$, $np\sigma $, $ns\sigma $ and $nd\sigma $
states. The symbols $\pm 1$ imply two opposite-sense electronic angular momenta
associated with the different signs of $\lambda $. Linear combinations of the $n\pi^{\pm 1}$ states give states  $np\pi '$ and $np\pi ''$, symmetric and antisymmetric with respect to
reflection in the plane containing the molecular axis. The other three states are symmetric. These states were established in Ref. \cite{larson05} to be the most important for DR, since they exhibit the largest
dependence on the vibrational coordinates. Therefore, $H_{int}+T_e$ assumes a
block-diagonal form with an infinite number of $5\times 5$ blocks corresponding
to $n=2,3,\cdots ,\infty ,$ in addition to the continuum. Since the only off-diagonal
couplings included in our analysis are those among the three $np$ states,
for simplicity we specify only these states in the formulas below.

The $np$-block of $H_{int}+T_e$ in the basis of the $np\pi ^{-1}$, $np\sigma $
and $\ np\pi ^{+1}$ states has the form \cite{koppel81} 
\begin{equation}
H_{int}(\mathcal{Q})=\left( 
\begin{array}{ccc}
E_{\pi } & \delta e^{i\varphi } & \gamma e^{2i\varphi } \\ 
\delta e^{-i\varphi } & E_{\sigma } & \delta e^{i\varphi } \\ 
\gamma e^{-2i\varphi } & \delta e^{-i\varphi } & E_{\pi }
\end{array}
\right) \,,  
\label{eq:Hint}
\end{equation}
where $E_{\sigma }$ and $E_{\pi }$ are the electronic energies of the $%
np\sigma $ and $np\pi ^{\pm 1}$ states at the linear ionic configuration; $%
\delta $ and $\gamma $ are the real, non-Born-Oppenheimer coupling elements.
We denote both the ($n\pi - n\pi $) and ($n\pi - n\sigma $) couplings as
Renner-Teller (RT) couplings, whereas in some previous studies, only the
former is denoted by this term. The couplings $\delta $ and $\gamma $
depend on $R_{CH},\ R_{CO}$, and $\theta $. The diagonalization of $[T_e+H_{int}](%
\mathcal{Q})$ is accomplished by the unitary transformation matrix $U$ \cite{koppel81}: 
\begin{equation}
U=\frac{1}{\sqrt{2}}\left( 
\begin{array}{ccc}
e^{i\varphi } & e^{i\varphi }w_{-} & e^{i\varphi }w_{+} \\ 
0 & \sqrt{2}w_{+} & -\sqrt{2}w_{-} \\ 
-e^{-i\varphi } & e^{-i\varphi }w_{-} & e^{-i\varphi }w_{+}
\end{array}
\right) \ ,  \label{eq:U_trans}
\end{equation}
with the abbreviations 
\begin{eqnarray}
\label{eq:w}
w_{\pm } &=&\sqrt{(1\pm \Delta /w)/2};\ \Delta =(E_{\sigma }-E_{\pi
}-\gamma)/2;  \notag \\
w &=&\sqrt{\Delta ^{2}+2\delta ^{2}}\,.
\end{eqnarray}
When diagonalized, the Hamiltonian becomes 
\begin{equation}
\label{eq:U}
U^{\dagger}[H_{int}+T_e]U=\mathrm{{diag}\{V_{\pi ''},V_{\sigma
},V_{\pi '}\}\,,}
\end{equation}
where 
\begin{eqnarray}
\label{eq:Ve}
V_{\pi ''} &=&E_{\pi }-\gamma \,,\quad V_{\sigma }=(E_{\sigma
}+E_{\pi }+\gamma )/2+w\ ,  \notag  \label{eq:21c} \\
V_{\pi '} &=&(E_{\sigma }+E_{\pi }+\gamma )/2-w.
\end{eqnarray}
The adiabatic potential energy surfaces $V_{\pi ', \pi'' ,\sigma }(\mathcal{Q}%
) $ are known from \textit{ab initio} calculations (see Ref. \cite{larson05}
for a detailed description). From Eq. (\ref{eq:21c})  we obtain
\begin{equation}
\label{eq:RT_param}
\gamma =E_{\pi }-V_{\pi ''}\,,\quad w=(V_{\sigma }-V_{\pi '})/2\,.
\end{equation}
Therefore,  the matrices $U$ in Eq. (\ref{eq:U_trans}) and $H_{int}$ in Eq. (\ref {eq:Hint}) are obtained from $V_{\pi',\pi'',\sigma }(\mathcal{Q})$.

\section{Reaction matrix and cross section for dissociative recombination}
\label{sec:cs}

As in our previous DR studies \cite{kokoouline01,kokoouline03a,kokoouline03b}, we employ multichannel quantum
defect theory and need to construct the reaction matrix $K$, which is related to the
potential as in Ref. \cite{SuzorWeiner98}. (We note, however, that the reaction matrix $K$ of Ref. \cite{SuzorWeiner98} is not the $K$-matrix of MQDT, which includes an additional factor of $-\pi$.)  First we introduce the diagonal quantum defect
matrix $\underline{\mu}$, whose nonvanishing elements $\mu _{i}$ are $V_{i}=-1/[2(n-\mu _{i})^{2}]$,
where $i$ is the electronic state index. Therefore, the diagonal form of the reaction matrix is directly obtained from the {\it ab initio} calculation.

\begin{figure}[h]
\includegraphics[width=30pc]{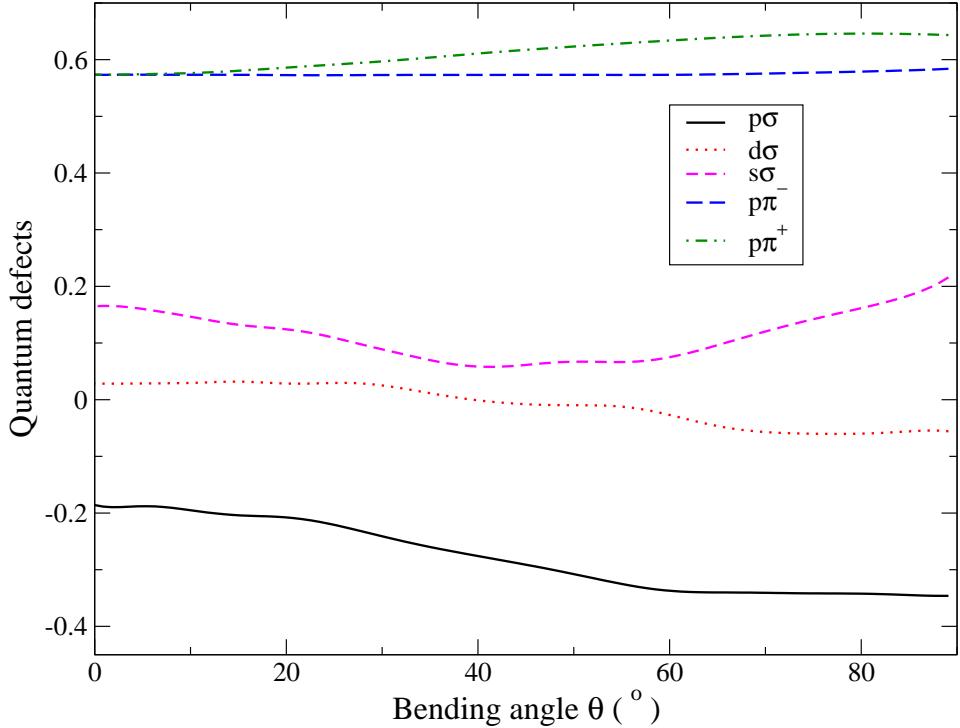}
\caption{(Color online) The quantum defects of the
electronic states included in the present study as a function of the bending
angle $\protect\theta$ for fixed $R_{CH}=2.0$ a.u. and $%
R_{CO}=2.0877$ a.u. Labeling the states with 
electronic momentum projection $\protect\lambda$ is an approximation and for large values of 
$\protect\theta$ it is not appropriate. However, the four $s\protect\sigma$, 
$p\protect\sigma$, $d\protect\sigma$, and $p\protect\pi'$ states are always
uncoupled from $p\protect\pi''$ states for nonlinear geometries. In the absence of the 
non-diagonal matrix elements in Eq. \ref{eq:Hint}  the $p\pi'$ and $p\pi''$ states would have the same quantum defects.
The Renner-Teller parameters are completely determined by the geometry dependence of $p\pi$ and $p\sigma$ defects (Eq. \ref{eq:RT_param}).}
\label{QD}
\end{figure}

The coupling $\delta$ in Eq. \ref{eq:Hint} is linear with $\theta$ for small $\theta$. Correspondingly, the splitting between adiabatic electronic energies $V_{\pi''}$ and $V_{\pi'}$  is quadratic with respect to $\theta$, and, therefore, the splitting between quantum defects $\mu_{\pi''}$ and $\mu_{\pi'}$ is also quadratic. This can be seen in Fig. \ref{QD}, where $\mu_{\pi''}$ is almost constant because it has a different symmetry and is unchanged due to the Renner-Teller effect. The quantum defect $\mu_{\pi'}$ is approximately quadratic due to the Renner-Teller effect until $\theta\approx 50^o$, beyond which the linear approximation for $\delta(\theta)$ is not adequate. In a preliminary calculation we used such a linear dependence of $\delta(\theta)$, because we have found that we can use formulas Eqs. \ref{eq:U_trans}, \ref{eq:w}, \ref{eq:U}, \ref{eq:Ve}, and \ref{eq:RT_param} and fit the model Hamiltonian accurately to the {\it ab initio} energies without the additional approximation of linearity of $\delta(\theta)$. The result of the preliminary calculation is consistent with the present calculation using the accurate fit to the {\it ab initio} energies.

The next step is to bring the reaction matrix into the representation of the Hamiltonian of Eq. \ref{eq:Hint}. Since the transformation matrix $U$ is known from Eq. \ref{eq:U}, the $K$ matrix in that representation reads 
\begin{equation}
K=U\,\mathrm{tan}(\pi \hat{\mu})\,U^{\dagger}\ .
\end{equation}
The matrices $\hat\mu$,\ $H_{int}+T_e,\ K,\ U$ are diagonal
for the  $ns\sigma$ and $nd\sigma$ states. We used quantum defects from
Ref. \cite{larson05} with $n=4$ for 
$ns\sigma$ and $n=3$ for the other states. Figure \ref{QD} gives
the quantum defects as functions of $\theta$, with $%
R_{CH}$ and $R_{CO}$ fixed at the ionic equilibrium values.

\begin{figure}[h]
\includegraphics[width=30pc]{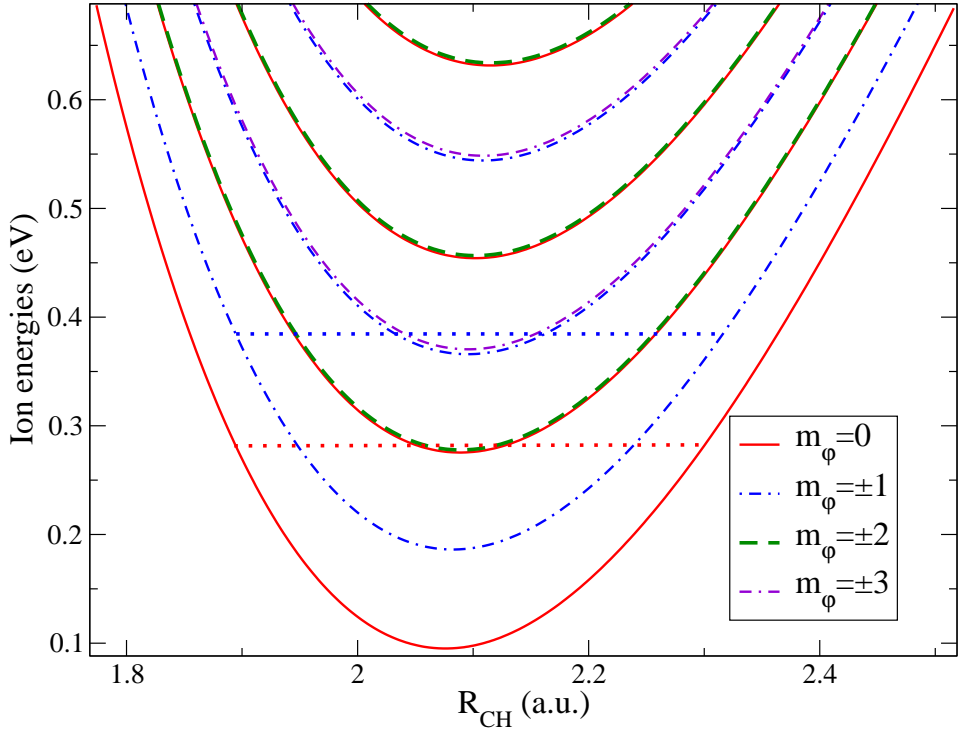}
\caption{(Color online) Several adiabatic potential curves for the HCO$^+$ ion as a function of the $R_{CH}$ distance for
different projections $m_\protect\protect\varphi$ of the vibrational angular momentum. The curves of the same color correspond to the vibrational states with the same quantum number $m_\varphi$ but different $L$. The energies of the lowest vibrational states with $m_\protect\protect\varphi=0$ and $m_\protect\protect\varphi=\pm 1$ are shown by two horizontal lines.}
\label{Energies}
\end{figure}

Once the reaction matrix $K_{i,i^{\prime }}$ is obtained, the DR treatment
is along the lines of Refs. \cite{kokoouline01,moshbach05} with the
following differences. We choose here the $R_{CH}$
distance as the dissociative adiabatic coordinate. This assumption is justified
because experimentally the H+CO channel is largely dominant at low
energies.
It has been long known that the CH+O channel is endothermic by 0.17 eV and
could have an energy barrier, while the OH product is not observed (see Ref.
\cite{lepadellec97} and references therein). Recently, branching ratios have
been measured \cite{Geppert:FD04} for the
DCO$^+$ dissociative recombination, confirming that the D+CO channel is by far
the dominant, with a ratio of 0.88. 
We keep the $%
R_{CO}$ distance frozen at its ionic equilibrium value. This assumption was made to simplify the treatment but it could result in an underestimation of the DR cross section. In order to account for the CO vibration one might use the hyper-spherical coordinates similar to Ref. \cite{larson05}, or else normal mode coordinates as in McCurdy {\it et al.} \cite{mccurdy03}. Although the CO bond is frozen, the Renner-Teller coupling physics is included in our model; we show below that it does increase the DR cross section significantly. We will briefly address below the possible influence of the CO vibration.

 Consequently, for every $R_{CH}$ distance we calculate matrix elements of the ``large'' reaction matrix $%
\mathcal{K}(R_{CH})$ 
\begin{equation}
\mathcal{K}_{j,j^{\prime }}(R_{CH})=\langle \Phi _{m_\varphi,L}|K_{i,i'}(\mathcal{Q})|\Phi _{m' _\varphi,L'}\rangle
_{\theta ,\varphi },
\end{equation}
where $\Phi _{m_\varphi,L}(R_{CH};\theta ,\varphi )$ are vibrational wavefunctions of
HCO$^{+}$ parametrically dependent on $R_{CH}$, while $R_{CO}$ is frozen. They are 
eigenfunctions of $H_{ion}$ with eigenvalues  $U_{m_{\varphi },L}^{+}(R_{CH})$.
Each index $j$ in $\mathcal{K}$ can be represented as $\{i,m_{\varphi },L\}$%
, where $m_{\varphi }$ specifies the projection of the vibrational angular momentum on the molecular axis, while the index $L$ distinguishes different vibrational states with the same 
$m_{\varphi }$. 
The rotation of the whole molecule is not considered. Fig. \ref{Energies} shows the vibrational eigenenergies $U_{m_{\varphi },L}^{+}$ as a function of $R_{CH}$.

It is informative to compare the vibrational energies obtained in our simplified approach with the exact calculation accounting for all four vibrational coordinates {puzzarini96}. We cannot compare the absolute values of the energies because the CO stretch is frozen, however, we can compare the energies of a few vibrational levels $\{v_1v_2^{l},v_3\}$ with respect to the energy $E=0$ of the ground vibrational level  $\{00^00\}$. Table \ref{tab:vibr_ene} compares the vibrational energies obtained in this study with accurate results from Ref. \cite{puzzarini96}. The overall accuracy of the present calculated vibrational levels is about 12 \%  or less. Since the vibrational wave function error is of order $\sqrt{0.12} \approx 0.35$, a conservative estimate of the error in our calculations of the final cross section would be of order $70\%$. 
\begin{table}[tbp]
\vspace{0.3cm}
\begin{tabular}{|p{2.cm}|p{4cm}|p{4.cm}|}
\hline
 $\{v_1v_2^{l},v_3\}$ & Present calculation & Puzzarini {\it et al.}\cite{puzzarini96} \\ \hline\hline
$10^00$& 2933 &  3090.40 \\
$01^10$& 740 &  830.7 \\
$02^00$& 1460 &  1641.14 \\
$03^10$& 2200 &  2458.9 \\
$04^00$& 2925 &  3256.94 \\
\hline
\end{tabular}
\vspace{0.3cm}
\caption{The table demonstrates the accuracy of the vibrational energies obtained in the present study neglecting the CO vibration with the exact calculation from Ref. \cite{puzzarini96}. Since the CO bond is frozen, we don't provide energies for excited $v_3$ modes. The overall error is about 12 \%, which translates into about  25 \% for vibrational wave functions. The energies are  given in cm$^{-1}$.}
\label{tab:vibr_ene}
\end{table}

The matrix $\mathcal{K}(R_{CH})$ is used to obtain the potential curves $U_a(R_{CH})$  of HCO as described in Ref. \cite{kokoouline01}. The curves in general have non-zero autoionization widths $\Gamma_a(R_{CH})$. 

The potential curves  $U_a(R_{CH})$ and their autoionization widths $\Gamma_{a}(R_{CH})$ are then
used to calculate the DR cross section. The cross section is calculated in a
manner similar to the procedure described in Refs. \cite{kokoouline01,moshbach05}, with
some adaptations. Specifically, we start from Eq. (5.19) of Ref. \cite{omalley66} for the
cross section of the process of dissociative attachment of the electron
to a neutral molecule, which applies equally to the dissociative recombination 
process considered here.  The resulting approximation adapted to this situation 
reads (see Eq.6 of Ref.\cite{kokoouline01}):
\begin{equation}
\label{eq:DA}
 \sigma_{DA}=  \frac{2\pi^2}{k_o^2}
         \frac{\Gamma_{a}(R_{CH})}{|U'(R_{CH})|}|\chi^+_o(R_{CH})|^2 e^{-\rho(E)}.
\end{equation}
Here $e^{-\rho(E)} \le 1$ describes the survival probability, which could
be less than unity if the system has a substantial probability to autoionize before it
dissociates; $k_o$ is the asymptotic wave number of the electron incident on 
vibrational level $o$ of the target molecule.
The distance $R_{CH}$ is a dissociation coordinate understood to be evaluated at the 
Condon point, which depends on the total energy as well as the initial target 
vibrational level $o$; $\chi^+_o(R_{CH})$ is the initial
vibrational wave function of the ion. 
The above formula is appropriate for the capture into a resonant potential 
curve that is energetically open for direct adiabatic dissociation. If the corresponding 
resonant state is closed, i.e. bound with respect to dissociation, it requires 
modification. To this end, we adapt Eq.~(4.2) of Bardsley \cite{bardsley66} 
to our present situation involving indirect DR.  The indirect process proceeds 
via capture into a bound (typically Rydberg) state, which eventually predissociates.  
For a case involving a single incident electron partial wave (e.g. $p \pi'$, 
for definiteness) and a single ionic target state, the fixed-nuclei autoionization 
width of the resonance potential curve will be denoted $\Gamma_{a}(R_{CH})$.  
Once the vibrational motion in the resonance potential is quantized into a 
vibrational resonance level, with radial wavefunctions $\chi^{\rm res}(R_{CH})$,
it should be remembered that only a subset $o$ of the target vibrational levels 
${v}$ will be energetically open, at any given total energy $E$.  Each resulting 
quantized resonance acquires a partial autoionization width $\Gamma_{a,o'}$ into 
an open vibrational channel $o'$, which is given approximately by
\begin{equation}
 \Gamma_{a,o'}= |\int_0^\infty \chi_{o'}^+(R_{CH}) \sqrt{\Gamma_a(R_{CH})} 
\chi^{\rm res}(R_{CH}) dR_{CH}|^2\,.
\end{equation}
The sum of these is then the total resonance autoionization width (within 
this approximation, neglecting Rydberg level perturbations of the MQDT type, 
which could sometimes produce complex, non-isolated resonances),
\begin{equation}
 \Gamma_{a,{\rm tot}}=\sum_{o'} \Gamma_{a,o'}
\end{equation}
while the total linewidth $ \Gamma $ of this quantized resonance also 
includes its predissociation partial linewidth, $\Gamma_d$:
$ \Gamma=\Gamma_d + \Gamma_{a,{\rm tot}}$. 
In this notation, the contribution to the total DR cross section will be 
the following, if only the ground vibrational level $o$ of the target is 
populated:
\begin{equation}
 \sigma^{\rm res}(E) = \frac{2 \pi^2}{k_o^2} \frac{1}{2\pi}   
\frac{\Gamma_{a,o}\Gamma_d} {(E-E_{\rm res})^2+{1\over4}\Gamma^2}\,.
\end{equation}

Our model does not account for vibrational motion along the CO bond and, therefore, positions of resonances cannot be compared directly with the experiment. Moreover, the resolution in the available experiments is usually not sufficient to resolve individual resonances. This suggests that we should average the cross section over energy. Each energetically-closed ionic bound vibrational state $c$ generates a Rydberg series of resonances $\epsilon_{n_c}$, numbered with the effective principal quantum number $n_c$. The cross section averaged over the energy interval between two resonances with energies $\epsilon_n$ and $\epsilon_{n+1}$ is given by
 \begin{equation}
  \langle\sigma(E)\rangle=\frac{1}{\Delta_n+\Delta_{n+1}}\int_{\epsilon_n-\Delta_n}^{\epsilon_n+\Delta_{n+1}}
   \sigma(E'){\rm d}E'\,, \quad 
   \Delta_{n+1}=\frac{\epsilon_{n+1}-\epsilon_n}2\,.
 \end{equation}
 Extending the limits of the integral to infinity, one can easily obtain the following:
   \begin{equation}
 \langle\sigma\rangle\approx\frac{2 \pi^2}{k_o^2} {\bigr (} \frac{\Gamma_{a,o}\Gamma_d}{\Gamma\Delta} {\bigr )}
%
 \end{equation}
where the $\Delta=\Delta_n+ \Delta_{n+1} \approx1/n_c^3$. Note that the quantity in parentheses here approaches a constant at sufficiently high effective principal quantum numbers $n_c$ in the relevant closed channel $c$, because each partial and total width in the parentheses of this formula should become proportional to $\Delta $ in this limit.  The total cross section is then calculated by summing up the average contribution from all Rydberg states (a sum over closed ionic channels).  In the limit where $\Gamma_d >> \Gamma_{a,{\rm tot}}$, this gives
\begin{equation}
\label{eq:sum_over_channels}
\langle\sigma\rangle=\frac{2\pi^2}{k_o^2}\sum_c|\langle\chi^{\rm res}(R_{CH}) | \sqrt{\Gamma_{a}(R_{CH})}|\chi^+_o(R_{CH})\rangle|^2 n_c^3  .
\end{equation}

The projection $M=m_\varphi+\lambda$ of the total angular momentum on the molecular axis O--C is a conserved quantity in our model. We calculate all the resonances  and the cross section for all allowed values of $M$. Since we are considering only $s\sigma,p\sigma,d\sigma,p\pi',\pi''$ quantum defects and the initial ion is in its ground state $m_\varphi=0$, the only possible values of $M$ for the final state are $0$ and $\pm1$. Resonance energies and widths are of course dependent only on $|M|$. Finally, we obtain for the cross section:
\begin{equation}
  \langle\sigma^{total}\rangle=\langle\sigma^{M=0}\rangle + 2\langle\sigma^{M=1}\rangle\,.
\end{equation}

\section{Results and discussion}
\label{sec:results}

\begin{figure}[h]
\includegraphics[width=30pc]{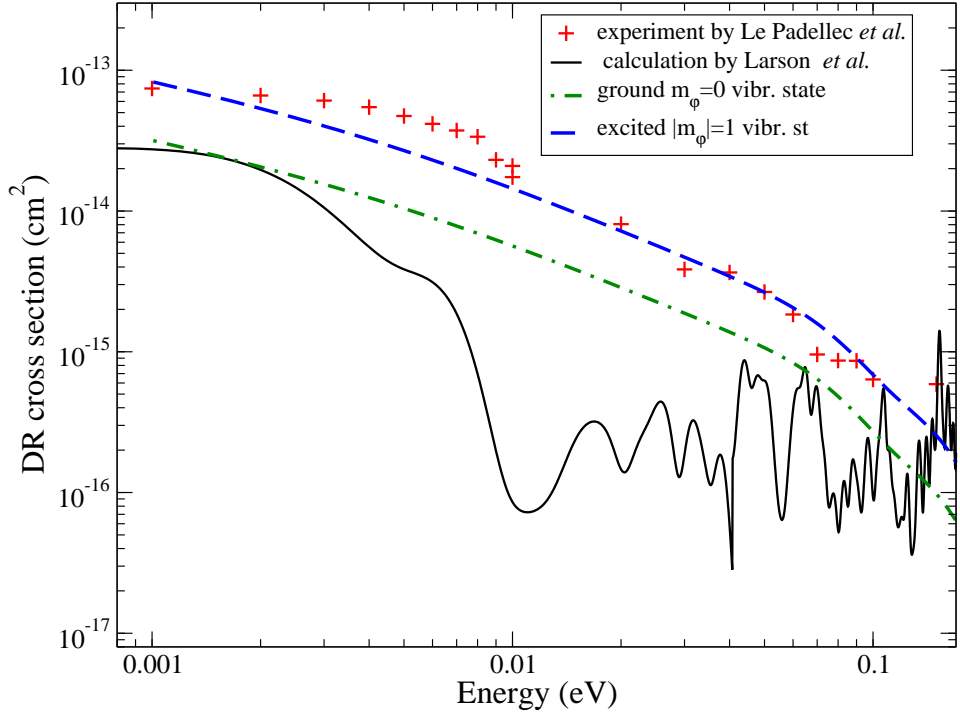}
\caption{(Color online) The figure shows the present
theoretical DR cross section (dashed and dot-dashed lines) for HCO$^+$ as a
function of  the incident energy $E$. The initial vibrational state is the ionic
ground state for the dot-dashed curve and the first excited state for the
dashed curve. The experimental \protect\cite{lepadellec97} (cross symbols)
and previous theoretical \protect\cite{larson05} (thin solid line) cross
sections are also shown for comparison. The theoretical curves include
the averaging over the electron energy distribution according to the
procedure described by Eq. (2) of Ref. \protect\cite{moshbach05} with $%
\Delta E_{\perp}=\Delta E_{||}=3$ meV. }
\label{cs}
\end{figure}

We have calculated DR cross sections 
for two different initial vibrational states $\chi^+_v(R_{CH})$ of the
ion: the ground state $v=0$ (which can be viewed as $\{00^00\}$) and the first excited 
state $v=1$ (which can be viewed as $\{01^10\}$) whose energies are represented  in Fig. \ref{Energies} 
with horizontal dotted lines. The energy difference between
the two states is about 0.1 eV. Figure \ref{cs} shows the two cross sections
with dot-dashed ($v=0$) and dashed ($v=1$) lines. 
Fig. \ref{cs} shows that the DR cross section depends strongly on the initial vibrational 
state. A similar strong dependence on the initial vibrational state 
has also been observed in experiments 
with other molecular ions. Assuming that the the initial experimental 
vibrational state distribution is in thermodynamic equilibrium, it is necessary to
average the cross section over the thermal distribution to compare with the experiment.
The first excited vibrational state has vibrational symmetry different from the
ground state. The vibrational angular momentum for the  states are $m_\varphi=$
1 and 0. Thus, the deexcitation process may be too slow to reach equilibrium,
resulting in a vibrational temperature that is higher than the electron temperature (300K in
the experiment of Ref. \cite{lepadellec97} ). The vibrational temperature in
the experiment of Le Padellec \textit{et al.} \cite{lepadellec97} is not
known. If it was in equilibrium with the electron temperature at 300K, the contribution from
the excited vibrational states would be small. But for a larger vibrational
temperature, for example 1000~K, the averaged cross section would be about a factor of
1.5 larger than the cross section for the ground vibrational state in Fig. \ref{cs}. The fact that in several
different experiments the measured DR rate ranges over values from $0.65-3\times10^{-7}$ cm$^{3}/$s at 300~K (see Ref.\cite{poterya05,adams84,amano90,ganguli88,gougousi97,laube98,leu73,rowe92,geppert05} 
and Fig. \ref{alpha} below) might conceivably derive from differences in the initial 
vibrational populations. Determination of the
actual experimental vibrational distribution and/or controlling it in HCO$^+$
could be an important step in understanding DR in small polyatomic ions. 

The present theoretical DR cross section is approximately a factor of 2 smaller 
than the experimental data. Assuming that this reflects a limitation of the present 
theoretical description this might derive from our approximation that freezes the CO bond length. 
However, the expected error of order 70\%
caused by our adiabatic approximation in the $R_{CH}$ coordinate should 
also be kept in mind when assessing the implications of this discrepancy.
Still, if the CO bond is allowed to vibrate, which makes the ion more floppy, 
the probabilities of capturing and predissociation will presumably be increased. 
Quantitatively, releasing the CO bond increases the density of HCO resonance states 
that can be populated in the electron-ion collision, and correspondingly, the sum in 
Eq. \ref{eq:sum_over_channels} is expected to increase. 

In afterglow plasma experiments, the measured observable is the DR rate 
thermally averaged over the kinetic energy distribution of colliding electrons and ions. 
This thermally averaged DR rate $\alpha(kT)$ is obtained from the DR cross section 
shown in  Fig. \ref{cs} according to Eq. (7) of Ref. \cite{kokoouline01}.  
The resulting theoretical rates (thermally averaged over the electron energies, 
for each of the two initial vibrational states) are shown 
in Fig. (\ref{alpha}) and compared with available experimental measurements.

 \begin{figure}[t]
\includegraphics[width=30pc]{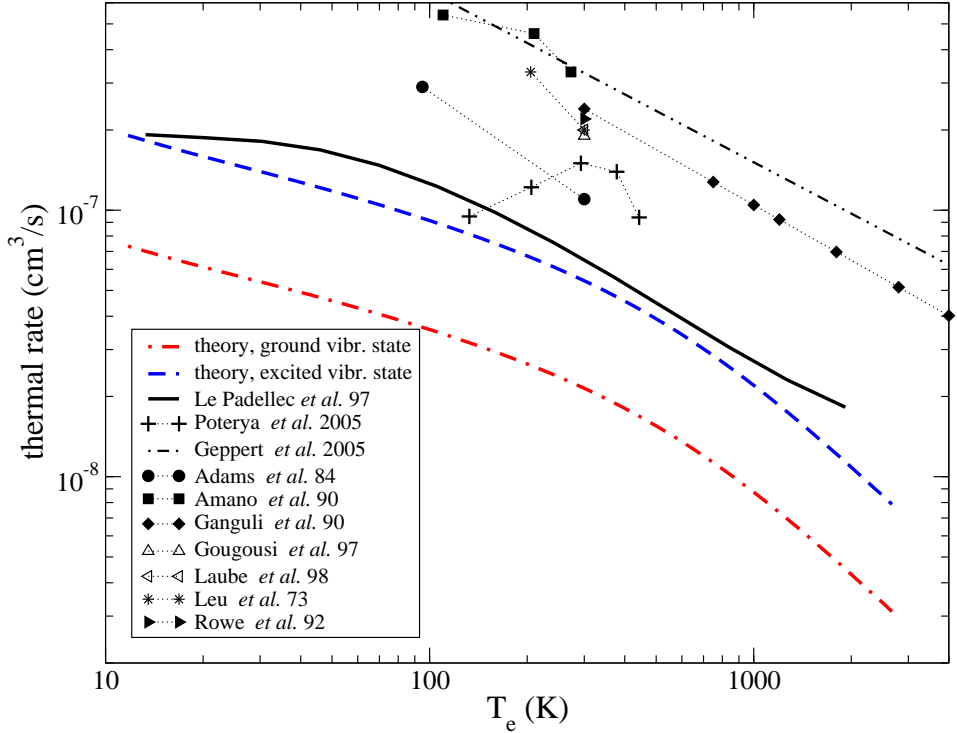}
\caption{(Color online) Theoretical and experimental DR thermal rates for HCO$^+$. 
The dot-dashed and dashed lines are the theoretical rates obtained for the ground 
and first excited vibrational states of the ion. The experimental rate (solid line) is 
obtained from the experimental cross section, Ref. \cite{lepadellec97}. 
The results from a number of other experiments with HCO$^+$ 
\cite{poterya05,adams84,amano90,ganguli88,gougousi97,laube98,
leu73,rowe92,geppert05} are also shown.}
\label{alpha}
\end{figure}

Although the theoretical DR cross section is still smaller by a 
factor 2 than the experimental results of Le
Padellec \textit{et al.} \cite{lepadellec97}, we find that inclusion of Renner-Teller 
coupling significantly increases the theoretical DR cross section. The previous 
theoretical study by Larson {\it et al.} \cite{larson05} that omitted 
Renner-Teller coupling, but which did account for the CO bond vibrations, gave a DR 
cross section significantly smaller than the present result. This is evidence 
for the important role of Renner-Teller coupling in the HCO$^+$ DR process.

Since the Born-Oppenheimer potential surfaces are the same for HCO (HCO$^+$) 
and DCO (DCO$^+$) we did a similar calculation for DR in DCO$^+$. The 
resulting DR cross sections for the ground and first excited vibrational level 
are very similar to the ones presented in Fig. \ref{cs} but smaller by a factor 
of 1.5. A similar dependence of the DR rate on the isotopologue masses was found 
in theory and experiment for the H$_3^+$, H$_2$D$^+$, D$_2$H$^+$, and 
D$_3^+$ ions \cite{kokoouline03b,kokoouline05}.

\section{Estimation of the effect of autoionization and CO vibration on the cross section}
\label{sec:auoionization}

\begin{figure}[h]
\vspace{5mm}
\includegraphics[width=30pc]{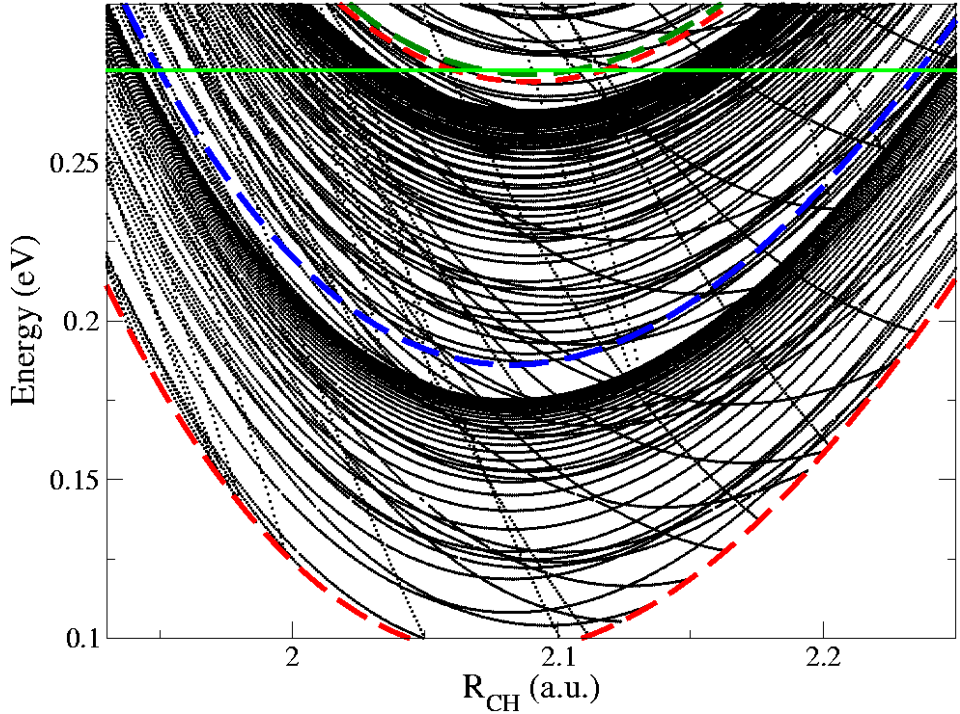}
\caption{(Color online) Resonance potential curves $U_a(R_{CH})$ of the neutral molecule 
states having magnetic quantum number $M=1$. The potential curves were calculated 
in the energy range where at least one electron-ion channel $U^+_i(R_{CH})$ (for the given 
$R_{CH}$) is open. Thus, the lower bound of the calculated curves corresponds 
to the ground potential curve $U^+_{0,0}(R_{CH})$ of the ion. The potential curves 
of the ion are also shown by dashed thick lines (see also Fig. \ref{Energies}). The 
Rydberg series are crowded just below the ionic potential curves. For clarity, visible gaps between 
the Rydberg series and the ionic curves have been intentionally introduced artificially, 
to show the behavior of the resonance curves belonging to higher ionic channels.   
The horizontal line represents the energy of the ground vibrational level. }
\label{resonances_m1}
\end{figure}

We should mention some approximations made in the present study that can
affect the theoretical DR cross section. First, our treatment as implemented here has not accounted
for the possible autoionization after the electron is captured by the ion,
before the neutral molecule has time to predissociate: in Eq. (\ref{eq:DA}),
the survival probability $e^{-\rho(E)}$ was set to be 1. The inclusion of
autoionization decreases the calculated DR cross section. The effect of
autoionization on the DR cross section can be estimated as following. Figure
\ref{resonances_m1} shows the calculated resonance curves $U_a(R_{CH})$ as
described above and in Ref. \cite{kokoouline01}. Above the lowest ionic curve
each of these potential curves has in general a non-zero width
$\Gamma_a(R_{CH})$, which represents the adiabatic autoionization rate. If there
is more than one open ionic channel $|j\rangle$ for a given $U_a(R_{CH})$, the
neutral molecule can decay into the $i$-th channel with a partial width
$\Gamma^{(i)}_{a}(R_{CH})$, where $\sum_i \Gamma^{(i)}_{a}(R_{CH})=\Gamma_a(R_{CH})$. 
Once the electron is captured, the system can evolve into two competing pathways: 
resulting in either autoionization or dissociation. The relative probabilities per unit of time 
for the two processes provide an estimate for the survival probability $e^{-\rho(E)}$. 

The largest partial autoionization widths are in general the widths of the resonance curves 
$U_a(R_{CH})$ belonging to Rydberg series associated with a few nearest closed ionic 
channels. Such resonance curves are very similar in shape to their parent ion potential 
curves, when the principal quantum number $n$ is high. Consequently, in the energy range 
of interest, these curves $U_a(R_{CH})$ are closed with respect to adiabatic dissociation. The 
dissociation can occur only through coupling to true dissociative states (i.e., the process denoted 
predissociation in molecular spectroscopy). The predissociation 
probability can be estimated using the Landau-Zener model. When the neutral molecule is 
vibrating along the $U_a(R_{CH})$ curve, every time it passes through an avoided 
crossing with another curve $U_{a'}(R_{CH})$, it can jump to the corresponding state 
$|a'\rangle$ via an adiabatic transition. We view such an adiabatic transition as the pathway 
to predissociation. The probability for adiabatic passage through an avoided 
crossing is given by the Landau-Zener formula \cite{LandauLifshitz}
\begin{equation}
\label{eq:P_adiab}
P_{a',a}=1-\exp\left(-\frac{2\pi \alpha^2}{\Delta F}\sqrt{\frac{m}{2\Delta E}}\right)\approx 
\frac{2\pi \alpha^2}{\Delta F}\sqrt{\frac{m}{2\Delta E}}\,
\end{equation}
where $\alpha$ is the non-diagonal coupling element between diabatic states, 
i.e. the states that would cross if the coupling was absent. Numerically, 
$\alpha$ is equal to half of the adiabatic potential curve splitting at the avoided crossing. $\Delta F$ is 
the absolute difference in slopes (net classical force) between crossing {\it diabatic} 
potential curves, $\Delta E$ is the classical kinetic energy at the avoided crossing, 
$m$ is the reduced mass of the system. The resonance curves shown in 
Fig. \ref{resonances_m1} demonstrate numerous avoided crossings.
\begin{table}[tbp]
\vspace{0.3cm}
\begin{tabular}{|p{2.cm}|p{2.3cm}|p{2.cm}|p{2.cm}|p{2cm}|p{2.5cm}|}
\hline
$R_{CH}$ a.u. & Energy (eV) & $\alpha$ (a.u.) & $\Delta F$ (a.u.) & $\Delta E$(a.u.) & $P_{a',a}$\\ \hline\hline
2.056 & 0.112 &  $1.18\times 10^{-5}$ & 0.0041 & 0.0059 & $0.9\times10^{-4}$\\
2.11 & 0.144 & $2.9\times 10^{-5}$& 0.0111 & 0.0047 & $2\times 10^{-4}$\\
1.93 & 0.34 & $4.0\times 10^{-5}$& 0.0036 & 1-37$\times10^{-4}$ & $14-83\times 10^{-4}$\\
2.02 & 0.283 & $1.0\times 10^{-5}$& 0.0022 & 1-37$\times10^{-4}$ & $1.2-7.2\times 10^{-4}$\\
\hline
\end{tabular}
\vspace{0.3cm}
\caption{The table demonstrates Landau-Zener parameters and probabilities for 
typical avoided crossings. The first two examples correspond to avoided 
crossings situated deeply below the energy of the ground vibrational level. 
The third and fourth examples corresponds to energies around and above 
the ground vibrational level. Thus, $\Delta E$ could be very different for the 
last two examples giving different probabilities. In general,
for any total energy of the system there are always some avoided crossings with 
small $\Delta E$  at the corresponding turning points.}
\label{tab:1}
\end{table}

Table \ref{tab:1} conveys an idea for the order of magnitude of the parameters in 
Eq. \ref{eq:P_adiab}. This table shows
two examples of avoided crossings that lie well below the energy at 
which the electron can be captured. At such large values of 
$\Delta E$, the Landau-Zener probability  
$~10^{-4}$, depends only weakly on the electron energy. The two other 
examples are taken from the region around the left turning point where the 
velocity of motion could be small or large depending on the energy of the 
electron. In cases like this one, we calculated the probability $P_{a',a}$ for a range of values of 
the kinetic energy $\Delta E$. There are many such avoided crossings that occur near left 
turning points and comparatively fewer at right turning points. 

During half ($\tau_{1/2}$) of an oscillation period, the 
system goes through $n_c$ avoided crossings. We estimate that $n_c$ is 
about 10 for a typical resonance curve. Therefore, the total predissociation 
probability $P_d$ is somewhere in the range 0.001-0.01 . In the above 
estimation of the predissociation probability we have assumed that only 
two states interact at each avoided crossing. However, in reality many of 
the avoided crossing cannot be considered as strictly two-state ones: three 
or more states $|a\rangle$ can participate. The probability of multistate 
crossing might conceivably give a larger predissociation probability.

The autoionization probability $P_{ion}$ during time $\tau_{1/2}$ is 
$P_{ion}\sim \tau_{1/2}\Gamma$, where $\Gamma$ is the total autoionization 
width. The largest $\Gamma$ is about $10^{-5}$ a.u., $\tau_{1/2}\sim \pi/\omega$, 
where $\omega/2$ is the frequency of oscillations in the ground vibrational level, 
$\hbar\omega/2=0.007$ a.u. Thus, $P_{ion}\sim \pi\Gamma/\omega\approx$ 0.002. 
This is presumably an upper bound on the autoionization probability. 

\begin{figure}[h]
\includegraphics[width=30pc]{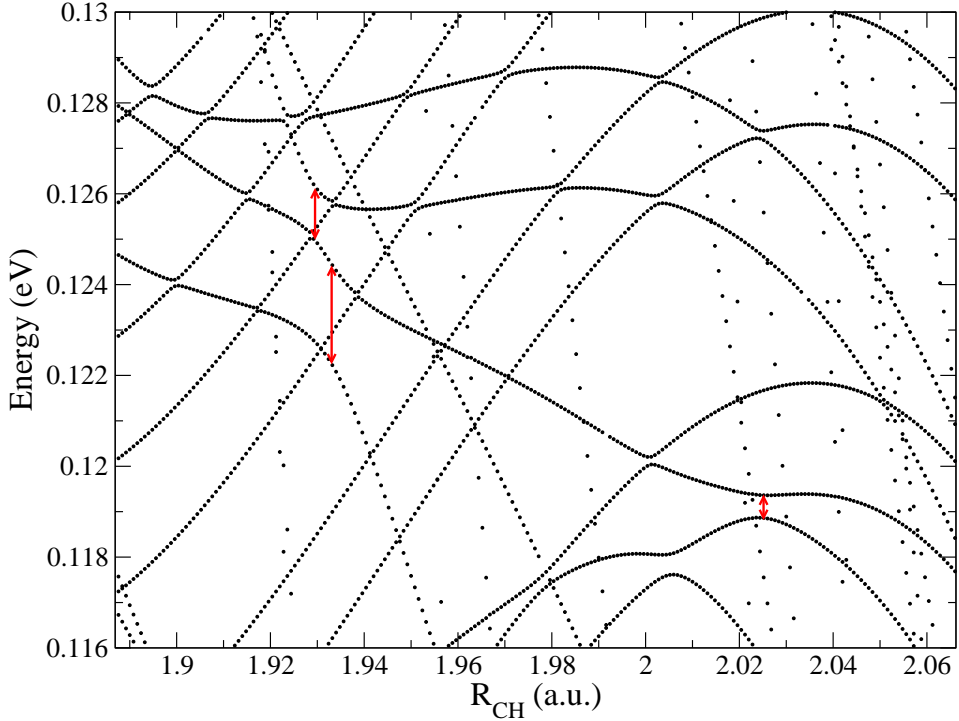}
\caption{(Color online) A part of the spectrum of Fig. \ref{resonances_m1} showing 
the interval of $R_{CH}$ corresponding to the left turning point of the vibrational 
motion. For convenience, the energy of the ground ionic curve $U^+_{0,0}(R_{CH})$ 
was subtracted from all the data. In the figure, curves with positive and small negative 
slopes correspond to Rydberg states converging to the nearest ionic thresholds. The 
probability to capture the electron tends to be high for such states, diminishing however 
at high principal quantum numbers as $n_c^{-3}$.}
\label{resonances_m1_2}
\end{figure}

As one can see, the autoionization probability could be competitive with
predissociation in our model if we take the smallest ratio $P_d/P_{ion}\sim
0.5$, but it is more likely that the ratio is of the order of 2-3 and
therefore, more favorable to predissociation. In the previous theoretical study
of DR in  H$_{3}^{+}$ \cite{kokoouline01}, the ratio was significantly larger.
An important difference with the H$_3^+$ study is the two degrees of freedom
taken into account when adiabatic potentials $U^+_i$ were calculated. As a
result, the density of resonance curves was much higher (factor 20-40). The
main contribution to the high density of resonance curves was from the states
with small principal quantum number corresponding to highly excited ionic
channels. This increased significantly the probability of predissociation. In
fact, in the case of HCO, the vibrating CO bond should also increase the
density of steep resonance curves in Figs. \ref{resonances_m1} and
\ref{resonances_m1_2}. This is because if the vibration along CO is quantized,
each of the curves shown in Fig. \ref{Energies} will produce a series additional ionic
curves with different CO quanta. Therefore, the number of ionic potential will be larger and, 
correspondingly, the density of resonance curves and avoided crossings in Figs. 
\ref{resonances_m1} and \ref{resonances_m1_2} will be increased too. 
This should increase the DR cross section 
(more states to be captured to) and the ratio $P_d/P_{ion}$ (more avoided crossings).

Another approximation is that the $n=2$ Rydberg states have dissociative 
character and are poorly described by this model.  However, we believe that the dominant DR pathways are triggered by an initial capture into higher Rydberg states, rather than being directly captured into $n=2$ states.

\section{Comparison with our previous theoretical study}
\label{sec:stefano_paper}

In the previous theoretical study of DR in HCO$^+$ 
published by some of us,\cite{larson05} the
Renner-Teller coupling was not taken into account because it appeared to have a
small effect on the potential surfaces of excited molecular states of the HCO
molecule. However, Renner-Teller coupling involves degenerate $\pi$ states
and therefore should play an important role in electron-ion scattering
especially when $p\pi$ and $p\sigma$ molecular potentials approach closely to
each other. Although Renner-Teller coupling was not included in the
previous study, some nonadiabatic effects were considered there.
The quantum defects $\mu({\cal Q})$ for $s\sigma,p\sigma,d\sigma$ states
obtained from {\it ab initio} calculation demonstrate sharp variations at some
configurations $\cal Q$ of the three HCO nuclei. Therefore, instead of using
the three adiabatic defects  $\mu({\cal Q})$, a numerical diabatization
procedure was applied to give a $\cal Q$-dependent 3x3 matrix $\mu^{(d)}({\cal
Q})$ of quantum defects. The $\cal Q$ dependence of the matrix is weaker and
the matrix has more information about nonadiabatic effects than the diagonal
matrix $\mu({\cal Q})$.  On the other hand the nonadiabatic effects
represented by non-diagonal elements of $\mu({\cal Q})$ might have a smaller
effect on the dynamics of the electron comparing with the nonadiabatic
couplings involving degenerate $\pi$ states. Indeed, the DR cross section
obtained in the previous study \cite{larson05} is much smaller than the
cross section of the present treatment. To verify that the increase in the DR
cross section is indeed due to the Renner-Teller coupling, we artificially set
the non-diagonal elements in the Hamiltonian of Eq. \ref{eq:Hint} to zero and
calculated again the cross section for $M=0$. 
The resulting DR cross section is about a
factor of three smaller than the cross section shown in Fig.~\ref{cs} and on
average it is close to the result of the previous study. It is worth mentioning
that the calculation of Ref.~\cite{larson05} accounts for all three vibrational
degrees of freedom but the very high vibrational levels that generate the steep
Rydberg $n=2$ curves in Fig. \ref{resonances_m1} (which have large widths and therefore lead
to an increase in the capture probability) were omitted; these two effects would tend to
balance each other, and might give roughly the same cross section as in the present
study when off-diagonal couplings are switched off. We speculate that if the CO 
bond vibration was included in the present treatment the effect of the 
Renner-Teller effect would likely be even more pronounced. 
To summarize our discussion of this point, we conclude that
the Renner-Teller effect appears to be the most important nonadiabatic 
coupling in the dissociative recombination of HCO$^+$. 

The theoretical cross section obtained in this study is inversely-proportional to
the incident energy below $0.1$ eV (see Fig. \ref{cs}).  Above 0.1 eV  the
cross section falls more rapidly with increasing energy. 
This is caused by the fact that at 0.1 eV
an additional ionic channel becomes open. 
That ionic channel is responsible for the large probability to capture the electron. 
Above that ionization threshold, the density of Rydberg states (in which capture
can occur) in much lower, as can be seen from Fig. \ref{resonances_m1}, and therefore
the capture probability drops.  Moreover, resonance states above that ionization threshold 
have another competing decay channel for autoionization, which steals flux away from 
the DR observable.
For the ground vibrational level $\{00^0v_3\}$ (dot-dashed line in Fig. \ref{cs} ), 
that additional open channel is $\{01^1v_3\}$, for the first excited level there are 
two nearby levels:  $\{02^0v_3\}$ and  $\{02^2v_3\}$. A similar behavior of the 
cross section was observed in DR of H$_3^+$ \cite{kokoouline01,
kokoouline03a,kokoouline03b}: the cross section drops down once a new 
ionic channel becomes open. The cross section from the previous theoretical  
study of HCO$^+$ (the solid line in Fig. \ref{cs}) does not show that behavior. 
It behaves smoothly as the energy
crosses the new ionization threshold. This is because the Renner-Teller effect was 
not accounted for:  hence transitions of the type $\{00^0v_3\}\longleftrightarrow \{01^1v_3\}$ 
are forbidden by the different symmetries of those vibrational states.

Another difference with the previous study is the manner in which the final cross section 
is calculated. In Ref. \cite{larson05}, the scattering matrix $S_{i,i'}$ for 
electron-ion collisions was explicitly calculated, but only in the electron-ion subset of 
channel space with no channel indices explicitly referring to dissociation. All of the channels $|i\rangle$ are 
vibronic states. However, some of the channels $|i\rangle$ are open for dissociation, and
as a result, the $S_{i,i'}$ matrix is not completely unitary. The defect from
unitarity was used to calculate the dissociative flux and the corresponding
DR cross section. The resulting cross section from this method has a rich structure with many 
Rydberg series of resonances. The cross section was then averaged over an appropriate 
thermal distribution of electron energies. The averaged cross section still has 
a number of resonances that are not seen in the comparatively low resolution experiment. In the 
present paper we have employed a different averaging procedure, as was discussed 
above.  This procedure gives a very smooth curve for the cross section.

\section{Conclusion}
\label{sec:conclusion}

In summary, we would like to stress the following:
\begin{itemize}
\item We have developed a theoretical approach to obtain the cross section 
for dissociative recombination of HCO$^+$. The approach is based on a model 
reaction matrix $K$ that includes the non-Born-Oppenheimer Renner-Teller 
coupling between doubly-degenerate vibrational modes of HCO$^+$ and 
the degenerate continuum states of the incident electron. 
The structure of the matrix $K$ is derived directly 
from the Hamiltonian that accounts for the Renner-Teller physics, with elements of the 
constructed $K$-matrix, if diagonalized, that reproduce the quantum defects 
obtained from {\it ab initio} calculations.

\item We have shown that even a modest Renner-Teller splitting can cause a
greatly enhanced DR rate. The previous theory without the Renner-Teller
coupling gave a lower DR cross section. The present DR cross sections are still
generally about a factor of 2 lower than the experimental cross sections.  

\item The vibration along the CO bond was not included in the present model. 
We believe that the inclusion of the CO bond should increase the density of 
HCO states available to be populated during the electron-ion collision. This 
will probably increase the DR cross section, but the quantitative amount 
remains to be determined by future studies. Another error expected to be of 
order 70\% is introduced by our use of an adiabatic approximation in the $R_{CH}$
coordinate, and this should also be improved in future calculations.

\item In the present treatment the dissociation occurs predominantly through the indirect 
pathway, i.e. through Rydberg states of HCO. The estimated autoionization probability 
is comparable with the predissociation probability, but it seems likely that inclusion of 
CO vibrations may increase the probability of predissociation.

\item The approach is quite general and can be applied to other closed-shell linear molecular 
ions, such as HCS$^+$.
\end{itemize}

This work has been supported by the National Science Foundation under Grant
No. PHY-0427460 and Grant No. PHY-0427376, by an allocation of NERSC
supercomputing resources and by an allocation of NCSA supercomputing
resources (project No. PHY040022). {\AA}. L. acknowledges support from the 
Swedish Research Council and the G\"{o}ran Gustafsson Foundation.


\begin{thebibliography}{99}

\bibitem{larsson00} M.~Larsson, in Adv. Ser. Phys. Chem. vol 10: {\it Photoionization and
 photodetachment} p. 693 ed C.~Y.~Ng (World Scientific, Singapore
 2000).

\bibitem{hickman87}  A.~P.~Hickman, J.Phys. B: At. Mol. Phys. \textbf{20}, 2091 (1987).
\bibitem{giusti80} A.~Giusti, J. Phys. B \textbf{13}, 3867 (1980).
\bibitem{giusti83}  A.~Giusti-Suzor, J.~N.~Bardsley, C.~Derkits, Phys. Rev. A  \textbf{28}, 682 (1983).
\bibitem{giusti84}   A.~Giusti-Suzor, Ch.~Jungen, J. Chem. Phys. \textbf{80},  986 (1984).
\bibitem{nakashima87}  K.~Nakashima, H.~Takagi, and H.~Nakamura, J. Chem. 
Phys. \textbf{86}, 726 (1987).
\bibitem{jungen97}  C.~Jungen, S.~C.~Ross, Phys. Rev. A \textbf{55}, R2503 (1997).
\bibitem{schneider97}  I.~F.~Schneider, C.~Stromholm, L.~Carata, X.~Urbain,
M.~Larsson, A.~Suzor-Weiner, J. Phys. B.: At. Mol. Opt. Phys. \textbf{30}, 2687 (1997). 

\bibitem{orel93}  A.~E.~Orel and K.~C.~Kulander, Phys. Rev. Lett. \textbf{71}%
, 4315 (1993). 

\bibitem{kokoouline01}  V.~Kokoouline, C.~H.~Greene, B.D.~Esry, Nature 
\textbf{412}, 891 (2001).

\bibitem{kokoouline03a}  V.~Kokoouline and C.~H.~Greene, Phys. Rev. Lett. 
\textbf{90}, 133201 (2003).

\bibitem{kokoouline03b}  V.~Kokoouline and C.~H.~Greene, Phys. Rev. A 
\textbf{68}, 012703 (2003).

\bibitem{LandauLifshitz}  L.~D.~Landau and E.~M.~Lifshitz, \textit{Quantum
Mechanics: Non-relativistic Theory} (2003) (Burlington MA: Butterworth
Heinemann) p. 409.

\bibitem{jahnteller} H.~A.~Jahn and E.~Teller, Proc. Roy. Soc. A {\bf 161}, 220 (1937).

\bibitem{longuet61}  H.~C.~Longuet-Higgins, in \textit{Advances in
Spectroscopy}, Interscience, New York, \textbf{II} 429 (1961).

\bibitem{ballhausen79} C.~J.~Ballhausen, J. Chem. Ed. {\bf 56}, 294 (1979).
(see also {\text http://www.quantum-chemistry-history.com/Ball\_Dat/QMInCom2.htm} ).

\bibitem{jungen80} Ch.~Jungen and A.~J.~Merer, Molec. Phys. {\bf 40}, 1 (1980).

\bibitem{duxbury98} G.~Duxbury, B.~D.~McDonald, M.~Van~Gogh, A.~Alijah, C.~Jungen, and H.~Palivan, J. Chem. Phys. {\bf 108}, 2336 (1998).

\bibitem{mccurdy03} C.~W.~McCurdy, W.~A.~Isaacs, H.~D.~Meyer, and T.~N.~Rescigno, Phys. Rev. A {\bf 67}, 042708 (2003).

\bibitem{herbst74} E.~Herbst and W.~Klemperer, Astr. J. {\bf 188}, 255 (1974) (and references therein).

\bibitem{lepadellec97}  A.~Le~Padellec, C.~Sheehan, D.~Talbi, and
J.~B.~A.~Mitchell, J. Phys. B: At. Mol. Opt. Phys. \textbf{30}, 319 (1997). 

\bibitem{poterya05}  V.~Poterya, J.~L.~McLain, N.~G.~Adams, and L.~M.~Babcock, J. Phys. Chem. \textbf{109}, 7181 (2005).

\bibitem{adams84} N.~G.~Adams, D.~Smith, and E.~Alge, J. Chem. Phys. {\bf 81}, 1778 (1984).

\bibitem{amano90}  T.~Amano, J. Chem. Phys. {\bf 92}, 6492 (1990).

\bibitem{ganguli88} B.~Ganguli, M.~A.~Biondi R.~Johnsen, and J.~L.~Dulaney, Phys. Rev. A {\bf 37}, 2543 (1988).

\bibitem{gougousi97} T.~Gougousi, M.~F.~Golde, and R.~Johnsen, Chem. Phys. Lett. {\bf 265}, 399 (1997).

\bibitem{laube98}  S.~Laube, A.~Le~Padellec, O.~Sidko, C.~Rebrion-Rowe, J.~B.~A.~Mitchell, and B.~R.~Rowe, J. Phys. B: At. Molec. Opt. Phys. {\bf 31}, 2111 (1998).

\bibitem{leu73} M.~T.~Leu, M.~A.~Biondi, and R.~Johnsen, Phys Rev. A {\bf 8}, 420 (1973).

\bibitem{rowe92}  B.~R.~Rowe, J.~C.~Gomet, A.~Canosa, C.~Rebrion, and J.~B.~A.~Mitchell, J. Chem. Phys. {\bf 96}, 1105 (1992).

\bibitem{geppert05} W.~Geppert, unpublished (2005).

\bibitem{larson05}  A.~Larson, S.~Tonzani, R.~Santra, and C.~H.~Greene, J.
Phys.: Confer. Ser. \textbf{4}, 148 (2005).

\bibitem{koppel81}  H.~Koppel, W.~Domcke, and L.~S.~Cederbaum, J. Chem. Phys. 
\textbf{74}, 2945 (1981).

\bibitem{SuzorWeiner98}  V.~Ngassam, A.~Florescu, L.~Pichl, I.~F.~Schneider, O.~Motapon, and A.~Suzor-Weiner, Eur. Phys. J. D \textbf{26} 165 (2003).

\bibitem{moshbach05}  V.~Kokoouline and C.~H.~Greene, J. Phys.: Confer. Ser. \textbf{4}, 74 (2005).

\bibitem{Geppert:FD04} W.~D.~Geppert, R.~Thomas, A.~Ehlerding,  J.~Semaniak, F.~Osterdahl, M.~af~Ugglas, N.~Djuric, A.~Paald, and M.~Larsson

\bibitem{puzzarini96} C.~Puzzarini, R.~Tarroni, P.~Palmieri, S.~Carter, and L.~Dore, Molec. Phys. {\bf 87}, 879 (1996).

\bibitem{omalley66} T.~F.~O'Malley, Phys. Rev. {\bf 150}, 14 (1966).

\bibitem{bardsley66} J.~N.~Bardsley, J. Phys. B {\bf 1}, 365 (1968).

\bibitem{kokoouline05}  V.~Kokoouline and C.~H.~Greene, Phys. Rev. A {\bf 72}, 022712 (2005).





\end{thebibliography}
\end{document}